# Method for 3D printing of cubic microbubbles: fully enclosed thin-walled microcavities with ultra-high aspect ratios

Sohail Khan[1], Zengbo Wang[1], Qingshan Yang[2], and Liyang Yue[1,*]

[1]School of Computer Science and Engineering, Bangor University, Dean Street, Bangor, Gwynedd, LL57 1UT, UK
[2]School of Metallurgy and Power Engineering, Chongqing University of Science and Technology, Chongqing, 401331, China

*Corresponding author email: l.yue@bangor.ac.uk

Keywords: two photon polymerisation; microbubble; thin-wall; high-aspect-ratio; micro 3D printing

**Abstract**
A microbubble is, in essence, a fully enclosed thin-walled microcavity. Unlike spherical microbubbles formed by expansions, 3D printing enables the free definition of their geometry, allowing precise control over shape and dimensions during fabrication. However, the geometric nature of microbubbles poses significant challenges for conventional photoresist-based lithographic microfabrication due to their fragile thin-walls, enclosed hollow volumes, and high sensitivity to mechanical stresses. These characteristics prevent developer solvents from accessing the internal cavities to remove unexposed photoresist. Two-photon polymerisation (2PP) is a laser-based 3D microprinting technique capable of sub-diffraction-limited resolution, offering exceptional design freedom for fabricating complex micro-architectures in photoresists. In this study, we demonstrate a 2PP-based method that overcomes these limitations and, for the first time, enables the successful fabrication of cubic microbubbles with ultra-high-aspect-ratio thin walls and fully enclosed microcavities using high-viscosity SU-8 2050 photoresist. The optimised process parameters and structural design facilitate efficient extraction of unexposed photoresist from the cavity interior while achieving a thin-wall ultra-high aspect ratio of approximately 340:1. The hollow nature and mechanical integrity of the printed structures were experimentally confirmed using micromanipulator-based probing. The proposed method maintains excellent dimensional accuracy and significantly reduces printing time for large-scale and high-count builds in 2PP processes. Such microbubbles are fundamental building blocks for optical resonators, microelectromechanical systems (MEMS) pressure sensors, microfluidic reaction chambers, and emerging metamaterials. This work establishes a robust parametric framework for their high-efficiency fabrication and provides valuable insights applicable to a broad range of complex microdevices.

## 1. Introduction

Two-photon polymerisation (2PP) has emerged as a transformative micro-/nanofabrication technology. Its first practical implementation, employing femtosecond (fs) lasers for 3D structuring, was demonstrated by Maruo et al. in 1997 [1, 2]. The process is initiated by two-photon absorption, a nonlinear optical phenomenon in which photoacid generator (PAG) molecules within the photoresist simultaneously absorb two near-infrared (NIR) photons from tightly focused fs pulses, typically at a wavelength of ~800 nm. The combined photon energy excites the PAG to an electronic state normally accessible only by a single ultraviolet (UV) photon, leading to its decomposition and the local generation of a strong Lewis acid confined to the laser focal volume (voxel). This highly localised reaction, which enables sub-diffraction-limited feature sizes through dose control [3, 4], triggers cationic ring-opening polymerisation of the epoxy-based photoresist, forming a densely crosslinked network that can be further reinforced by post-exposure baking [5]. When integrated with high-precision scanning equipment, such as piezo positioning stages and galvanometer scanners, a 2PP system enables the direct fabrication of complex freeform 3D architectures voxel-by-voxel and layer-by-layer, with applications in photonics, micro-optics, biomedical engineering, microrobotics, and microfluidics [6–10].

Despite these advancements, 2PP and other microfabrication techniques continue to encounter substantial challenges in producing certain geometries due to fundamental process limitations. Among these, microbubble – an ultra-high aspect ratio, thin-walled, fully enclosed microcavities remain some of the most difficult structures to realise. Firstly, their thin film and high-aspect-ratio features are prone to deformation or collapse during laser writing, development, or post-exposure processing in 2PP protocol because of the mechanical stresses caused by voxel elongation, diffusion-driven overcuring, heat accumulation, etc. [11–15]. Several strategies have been proposed in recent years to mitigate these issues. Kamranikia et al. demonstrated high-aspect-ratio polymer micropillars while addressing collapse and deformation caused by capillary forces [12]. Liu et al. reported the open 'fishnet' 3D architectures incorporating multilayer thin walls [16]. LaFratta et al. provided a comprehensive review of how photoresist properties influence the fabrication of mechanically fragile thin walls and internal features in 2PP [17].

Secondly, for microbubble and similar fully enclosed microcavities, the removal of unexposed photoresist from the internal volume represents the most critical process challenge. In conventional 2PP workflows, once a microcavity is completely sealed, the liquid developer cannot penetrate the enclosed volume to dissolve the trapped photoresist, often resulting in a solid core despite the presence of unexposed material [18]. This issue arises from the viscous nature of the photoresist, which prevents effective flush-through removal unless the cavity remains open and accessible to the development solvent [18]. Similar challenges have been reported in stereolithography, where retaining empty spaces within solidified regions is hindered by 'print-through' of excess polymerisation, an effect analogous to the trapping of unexposed photoresist in 2PP [19]. Therefore, deliberate design modifications that enable developer solvent access are essential for the reliable fabrication of microbubbles. Previous studies have only demonstrated the fabrication of open cavities using 2PP in optical resonators and micro-hollow fibres [20-22], and, to the best of our knowledge, no prior work has yet investigated the production of microbubbles and other fully enclosed microcavities using 2PP.

Besides, microbubbles also represent a type of geometries with significant potential to accelerate 2PP process, as only the outer "shell" of the structure must be fabricated in applications that do not require internal features. Rohbeck et al. demonstrated that the printing speed, feasibility, and material efficiency of 2PP can be substantially enhanced through optimisation of geometrical design [23]. SU-8 2050, an epoxy-based negative-tone high-viscosity photoresist (~4500–5500 mm²/s), is widely used for high-aspect-ratio microfabrication due to its excellent mechanical strength, chemical resistance, and capability to produce thick films (>50 μm) in photolithographic processes [24, 25]. Its high viscosity suppresses flow, deformation, and vibration under gravity or capillary forces during laser writing [14], thereby helping microbubbles maintain their shape prior to development. Compared with existing 2PP parallelisation strategies, which reduce fabrication time by increasing throughput using complex and costly optical systems [11, 26–28], the combination of microbubbles and high-viscosity photoresists offers a simpler and more efficient approach for achieving the same purpose.

In this work, we addressed the aforementioned process challenges and successfully developed a robust 2PP method for fabrication of cubic microbubbles using high-viscosity SU-8 2050 photoresist. The reason a cubic geometry was selected is to realise thin-film wall and achieve an ultra-high aspect ratio between the microbubble length/height and wall thickness, as it enables more precise photon dose control during long, constant-speed laser scans along the side length compared to other shapes. Laser writing parameters, scanning strategies, and post-processing procedures were optimised to overcome the key fabrication challenges identified above, thereby enabling the reliable replication of the targeted cubic microbubbles, as verified experimentally by confirmation of their hollow nature.

## 2. Method

Experiments were conducted in three phases. First, freestanding square thin-walls were fabricated to assess geometric limits and optimise process parameters, including maximum microbubble-wall aspect ratios (length/height-to-thickness, LH/T) realised in this approach. Second, a top-surface structure referred to as a 'roof' was fabricated to span the freestanding thin-walls to create cubic microbubble. An optimised 2PP protocol was developed to precisely manage photoresist shrinkage during laser direct writing, producing microchannel gaps at the interface between the polymerised resist and the substrate that allowed developer solvent ingress and venting during development. These gaps moderately self-seal during post-processing, as residual solvent or incomplete polymerisation causes slight structural expansion [29, 30]. Third, a piezo-driven micromanipulator was employed to mechanically tip the fabricated cubic microbubble, enabling exposure of their hollow interiors for morphological inspection.

The schematic of the optical setup of the 2PP platform used in this study are shown in Fig. 1(a). This custom-built system provides high versatility and sub-micron resolution for 2PP fabrication. The core of the setup is a Toptica FemtoFiber Smart fs laser with an average power of 50 mW, a wavelength of $\lambda$ = 780 nm, a pulse duration of $\tau \leq$ 100 fs, and a repetition rate of $f$ = 100 MHz. The laser output first passes through a shutter, which enables precise interruption of the beam during writing. An attenuator is placed downstream to finely regulate the incident laser power, allowing adjustment to match the polymerisation threshold of the photoresist and to achieve the required voxel size for microbubble fabrication. The beam then propagates through a beam expander to optimise the Gaussian beam profile and adjust the beam diameter for efficient filling of the back aperture of a plan fluorite oil-immersion objective (Olympus RMS100X-PFO, 100×, numerical aperture $NA$ = 1.3). The resolution of 2PP is dominated by the imaging resolution of the optical system. In particular, the estimated lateral ($r_{xy}$) and axial ($r_z$) dimensions of light absorption volume can be approximated by the following equations [31, 32]:

$$r_{xy} \approx \frac{\lambda}{2\sqrt{2}\,NA} \tag{1}$$

$$r_z \approx \frac{2\lambda n_p}{\sqrt{2}\,NA^2} \tag{2}$$

, in which $\lambda$ denotes light wavelength, $n_p$ represents the refractive index of photoresist, and $NA$ is the numerical aperture of the objective lens. In this case, the refractive index $n_p$ of the SU-8 2000 series is 1.5795 at the laser wavelength of 780 nm [33]. Under these conditions, the estimated lateral and axial resolutions ($r_{xy}$ and $r_z$) are approximately 212 nm and 1030 nm, respectively.

The substrate, typically a glass slide spin-coated with SU-8 2050 photoresist, was mounted on a high-precision piezo micro-positioning stage (Physik Instrumente – PI P-611.3 NanoCube with E-517 Controller). Immersion oil (Olympus Type F, $n_{oil}$ = 1.518) was applied between the objective front lens and the cover glass, which was placed on top of the photoresist-coated glass substrate to ensure stable optical coupling and to minimise spherical aberrations. A monochromatic camera was positioned above the objective to provide real-time process monitoring, accurate focal plane identification, and precise alignment. The picture of the complete experimental setup is illustrated in the inset of Fig. 1(a). The open-source 3D-printing software Cura® was used to slice the models and generate G-code for single-pass thin-wall scanning, as shown in Fig. 1(b) I. An in-house JAVA translator then converted the G-code into PI General Command Set (GCS) codes to drive the piezo stage and laser shutter, enabling layer-stacking scans for freestanding thin-walls and cubic microbubble, as shown in Fig. 1(b) II and III, respectively.

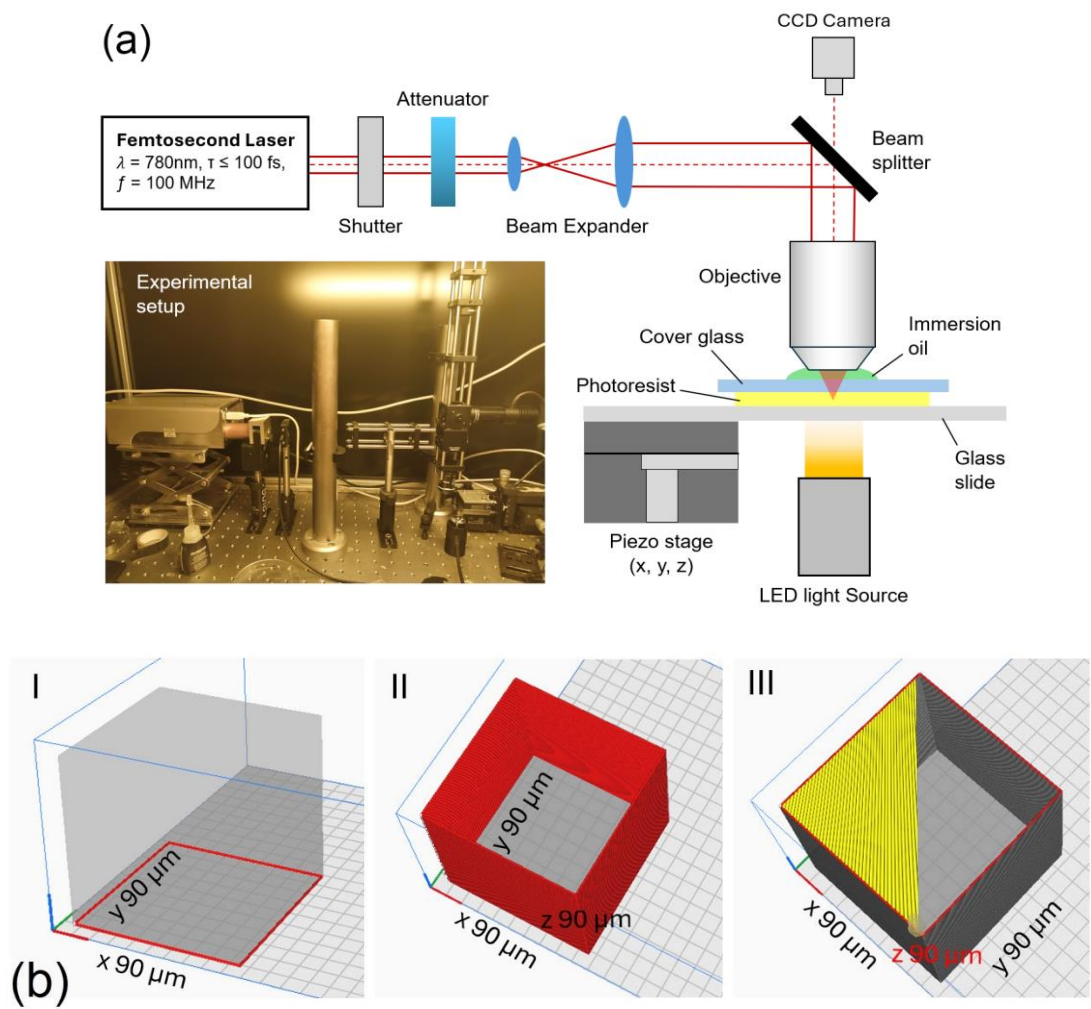

Fig. 1 (a) Schematic of the optical setup of the 2PP platform used in this study (b) 2PP slicing models for the structures in this study.

For sample preparation, standard glass slides were thoroughly cleaned by sequential ultrasonication in acetone, isopropanol (IPA), and deionized water, and subsequently dried under a nitrogen stream to ensure a pristine, contaminant-free surface. SU-8 2050 photoresist was then dispensed onto the cleaned substrates by drop casting and spin-coated at 1500 rpm for 45 seconds, producing a uniform film with a thickness of approximately 120 μm, in agreement with the rheological properties and manufacturer-provided spin curves for this photoresist viscosity [33]. A two-step soft-bake protocol was employed to remove residual solvent and promote film densification: the temperature was first ramped to 50 °C and held for 8 min, followed by heating to 98 °C for 14 minutes. After soft baking, the substrates were allowed to cool gradually to room temperature to minimise thermal stress and prevent microcrack formation. This procedure resulted in a uniform, defect-free, and chemically stabilised resist film suitable for subsequent 2PP laser writing. For development, the samples were post-exposure baked at 98 °C for 10 minutes and then cooled to room temperature prior to immersion in propylene glycol monomethyl ether acetate (PGMEA) development solvent for 10 minutes. The developed samples were subsequently rinsed in IPA to remove residual solvent, yielding well-defined and mechanically stable microstructures. All microstructures were uncoated and characterised using a Zeiss EVO 10 scanning electron microscope (SEM) equipped with an extended-range cascade current detector (C2DX) for high-resolution imaging.

## 3. Results and discussion
### 3.1 Fabrication of high-aspect-ratio thin-walls

Fabrication of freestanding, thin-walled cubic microstructures without roofs served as the first step toward producing cubic microbubbles with fully enclosed cavities. A single-pass printing strategy was implemented to minimise wall thickness, with each layer written at a uniform height of 800 nm, as shown in Fig. 1(b) (I), (II). The scanning speed was maintained at 100 μm/s for all experiments. A comprehensive study of the 2PP fabrication parameters was conducted, and the resulting wall thicknesses and LH/T aspect ratios obtained under varying laser output powers are presented in Fig. 2(a). This investigation identified a stable ultra-high aspect ratio regime for freestanding thin-walls produced at a laser power of approximately 4.1 mW (indicated by the black solid line in Fig. 2(a)). An example of this with a side length/height of 90 μm and a wall thickness of 267 nm is shown in Fig. 2(b), corresponding to an ultra-high LH/T aspect ratio of 337:1. It is notable that the inward deformation along the $x$-direction is evident in Fig. 2(b). Such features are commonly observed in high-aspect-ratio thin-walled freestanding structures fabricated without additional support. In contrast, this deformation effect can be significantly reduced by decreasing the LH/T aspect ratio, as demonstrated in Fig. 2(c), which shows a freestanding thin-walled structure with a 50-μm side length/height fabricated by the same 2PP parameters.

Similarly, localised deformations also occur at the bonding interface between the thin wall and the glass substrate, as revealed by angular observations of the rear side of the freestanding wall, as shown in Fig. 2(d). These deformations, wherever at the top of the thin wall as shown in Fig. 2(b) or at the glass interface as shown in Fig. 2(d), primarily arise from polymerisation shrinkage, which induces substantial lateral stresses at both the top and the base. Although the material is initially constrained by adhesion to the substrate along the laser scanning path, the generated stresses can exceed the interfacial bonding strength at several particular positions. This leads to partial detachment or lateral displacement, creating small gaps manifested as leaning or bending near the interface, as shown in the insets of Fig. 2(d). Solvent effects during development and drying introduce additional capillary and swelling stresses, further distorting these features [29, 34]. Interestingly, because the bottom filled layer of these structures was intentionally omitted in the design, the gaps formed at the bonding interface also provide microchannels through which developer solvent can enter the central fully enclosed cavity, facilitating the removal of unexposed photoresist, thereby offering a practical solution for fabricating microbubbles using 2PP.

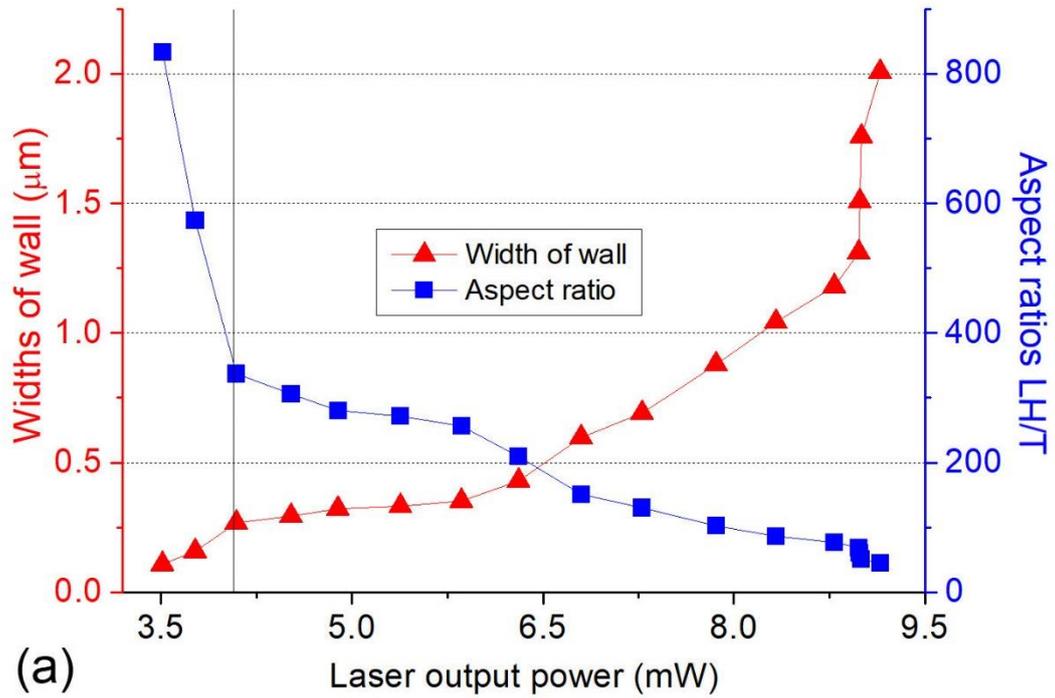

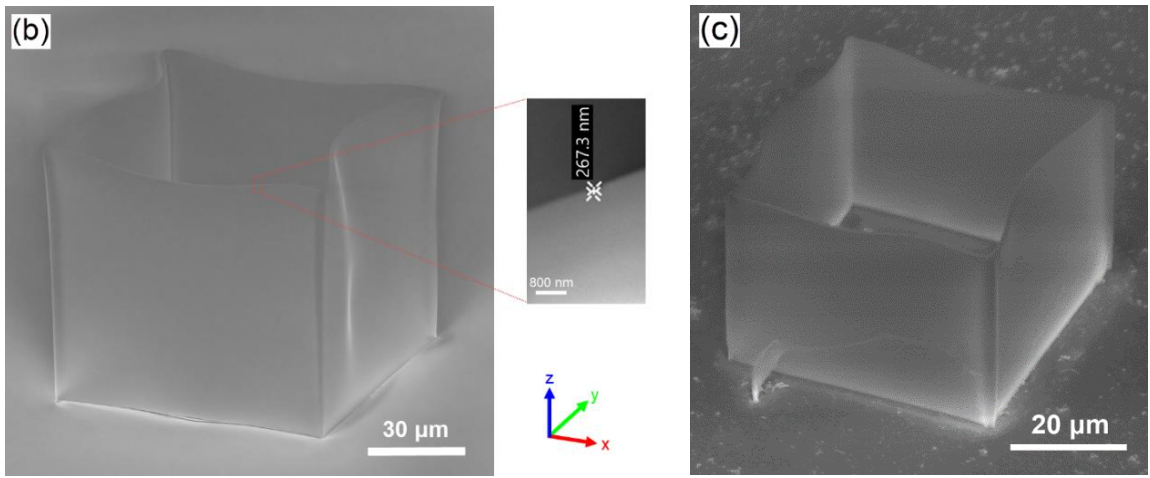

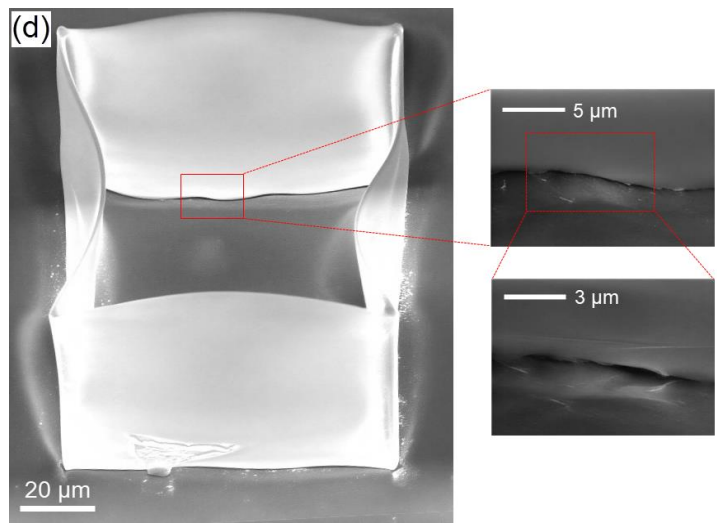

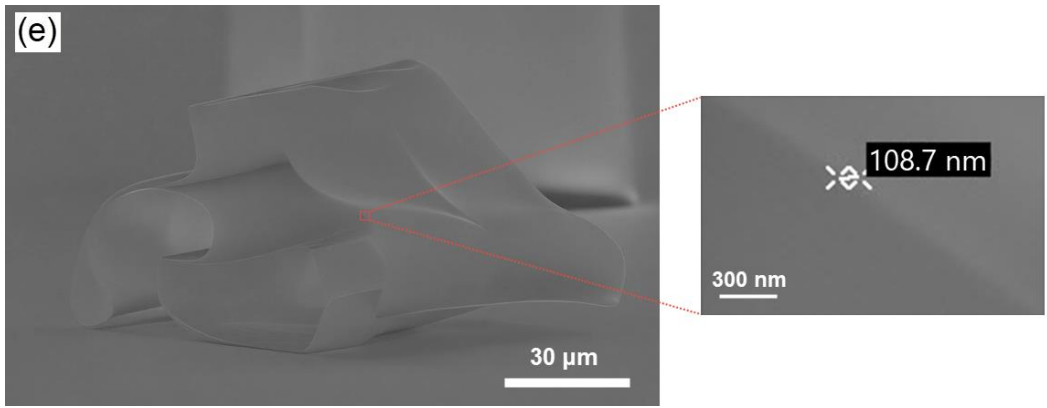

Fig. 2 (a) Wall thicknesses and aspect ratios of the fabricated thin-walled structures against laser output power. Freestanding cubic thin-walls with (b) 90-μm side length (c) 50-μm side length. (d) The deformation-induced gap (developer channel) formed at the bottom bonding interface. (e) The collapsed and folded thin-wall structures fabricated by low laser output power.

Meanwhile, further reductions in laser output power generated even thinner walls. At 3.6 mW, for example, a minimum wall thickness of approximately 109 nm was achieved. However, such ultrathin structures with aspect ratios exceeding 800 are mechanically fragile. The stresses generated during polymerisation and subsequent development surpass the adhesion strength of the SU-8 2050 photoresist to the glass substrate, leading to complete detachment and folding of the thin-wall, even though the walls themselves remain structurally intact as shown in Fig. 2(e).

### 3.2 Fabrication of cubic microbubbles

Building on prior studies of freestanding thin-wall microstructures, we further developed this method for the fabrication of hollow architectures, focusing on cubic microbubbles. In addition to their relevance for photonics and microfluidic applications, the proposed microbubbles minimise the volume of polymerised material, enabling substantially reduced laser scanning times compared with fully solid structures and representing an important advance in the optimisation of 2PP processes.

The deformation as shown in Fig. 2(b) is predominantly localised at the upper regions of the free-standing walls. To mitigate this critical bending, a top-surface roof was fabricated to span the side walls, enabling the formation of straight-cornered, flat-thin-walled cubic microbubbles and enhancing overall structural stability. The roof was realised by superimposing multiple polymerised layers to ensure sufficient mechanical rigidity and resistance to polymerisation shrinkage and capillary forces during development. Initial experiments focused on smaller cubic microbubbles with the 50-μm side length and varying heights. By increasing the roof thickness to 2 to 5 writing layers of 2PP, cubic microbubbles with straight corners and flat walls were successfully fabricated at a laser output power of 4.1 mW and a scanning speed of 100 μm/s, as shown in Fig. 3(a) and (b). These results demonstrate that a side-wall thickness of approximately 270 nm (Fig. 2(a)), combined with a multi-layer roof, provides sufficient structural reinforcement.

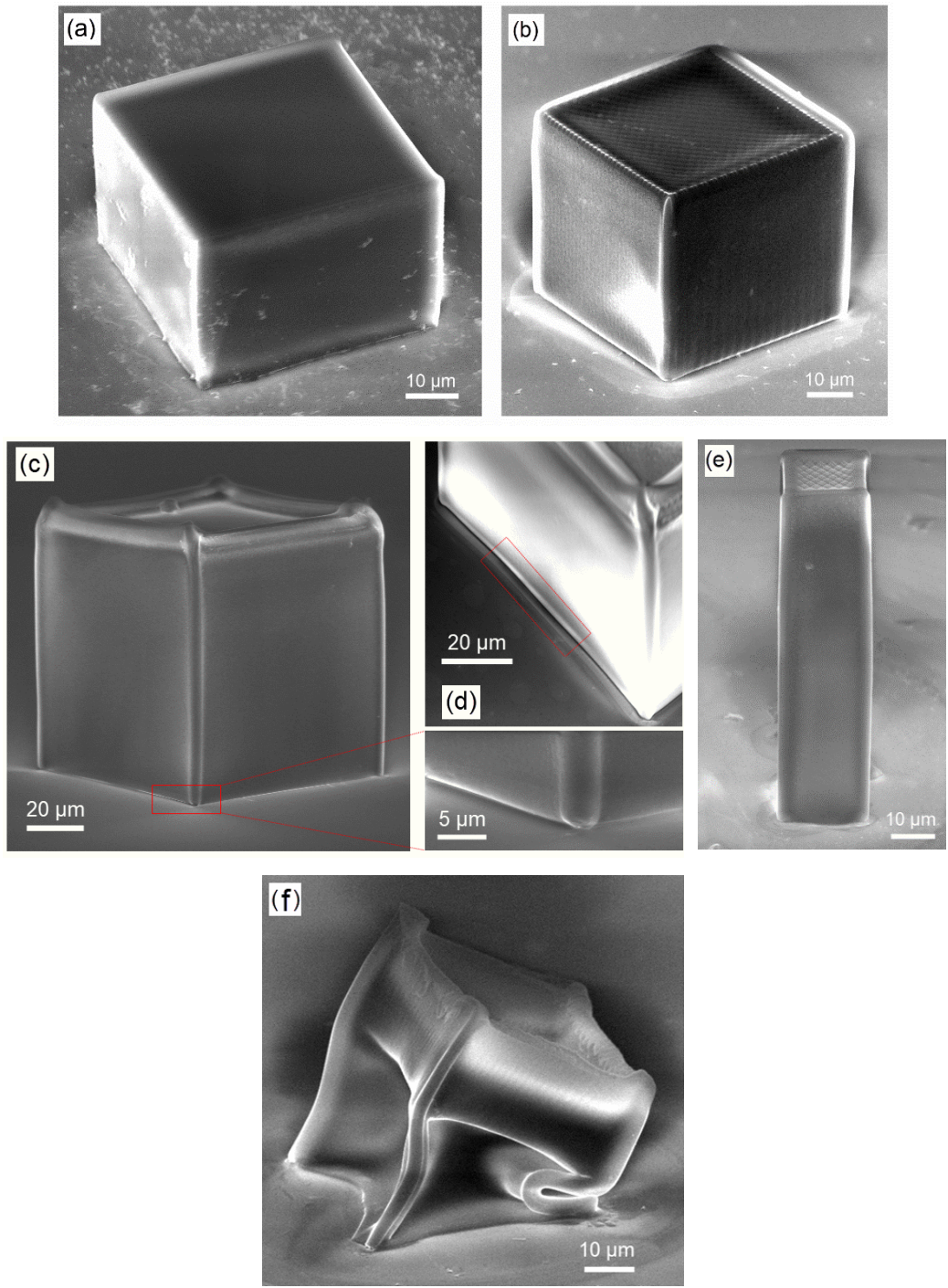

Fig. 3 (a) Cubic microbubble with a side length of 50 μm and a height of 30 μm. (b) Cubic microbubble with a side length and height of 50 μm. (c) Cubic microbubble with a side length and height of 90 μm; Inset shows a magnified corner. (d) An additional tilted-view image of the interface (e) Cubic microbubble with height-to-width (H/W) aspect ratios of 4.5:1 (f) Collapsed microbubble with roof hatching distance of 800 nm

This precise control of wall thickness and structural integrity also enabled the formation of microchannel gaps at the interface between the thin-wall and the glass substrate, which in turn the fabrication of cubic microbubble with higher LH/T aspect ratios, e.g. a side length/height of 90 μm using the same 2PP parameters, as illustrated in Fig. 3(c). These gaps, which facilitate the extraction of unexposed photoresist, moderately enclosed during post-processing and development due to slight structural expansion caused by residual solvent or incomplete polymerisation [29, 30]. The inset of Fig. 3(c) presents a magnified view of a 'sealed' microchannel gap at the corner. Similar features are also observed at other interfaces, as highlighted by the red box in an additional tilted-view image in Fig. 3(d). Compared with the microchannel gaps shown in Fig. 2(d), these sealed gaps are more uniform and continuous, appearing as a rounded edge bonded to the glass substrate and thereby forming an integrated enclosed microbubble. Using this established method, we further investigated the feasibility of fabricating cubic microbubble with high height-to-width (H/W) aspect ratios. An example achieving a maximum H/W aspect ratio of 4.5:1 was realised, as shown in Fig. 3(e). This result confirms the capability of the established method to realise fully enclosed, elongated cubic microbubbles with H/W aspect ratios greater than 1, highlighting its robustness in surpassing the structural and geometric limitations commonly encountered in microcavity applications.

Simultaneously, we investigated the influence of roof hatching distance on structural stability and material consolidation. Increasing the hatching distance generally reduces the degree of conversion, leading to lower Young's modulus and hardness of the cured photoresist [35], and thus a less dense and mechanically weaker structure. As shown in Fig. 3(f), despite employing 5 top layers for the roof, a large hatching distance of 800 nm resulted in structural collapse due to insufficient mechanical integrity. This inward collapse is attributed to the loosely formed roof caused by the hatching distance being significantly larger than the theoretical lateral resolution of the system (~212 nm), which created excessive open channels for development solvent penetration and IPA evaporation in the development phase. The resulting capillary forces, induced by surface tension during solvent evaporation, led to inward bending or fracture when the aspect ratio exceeded a critical threshold or the mechanical stiffness was inadequate. This inward collapse can be effectively mitigated by reducing the roof hatching distance to 500 nm, as demonstrated by the 5-layer enclosed cavity in Fig. 3(c). Further reduction of hatching distance to 200 nm enabled the fabrication of sound, straight-cornered cubic microbubbles using only two roof layers, as shown in Fig. 3(b). The decreased hatching distance produces a denser and more continuous polymer network within each layer, enhancing local mechanical strength and adhesion, as the cross-link density directly governs the stiffness and strength of the polymerised photoresist. This optimisation yields a robust and compact roof even with fewer layers, demonstrating that precise control of hatching parameters is critical for improving the mechanical performance and structural fidelity of 2PP-fabricated microbubble architectures.

### 3.3 Microcavity verification
An additional experiment was conducted to verify the hollow interior of the cubic microbubbles fabricated using the proposed method and to confirm the complete removal of unexposed photoresist from their interiors. A mobile piezo-actuated micromanipulator (miBot®, Imina Technologies), illustrated in Fig. 4(a), was employed to mechanically displace individual microcavities from the glass substrate. Under angular magnifying camera monitoring, a pointed probe mounted on the micromanipulator was carefully advanced to gently tap the microcavities, causing them to tip over, as shown in the diagram of Fig. 4(b). This manipulation enabled subsequent SEM observation of their internal architecture. The operation was technically challenging due to the extreme size contrast between the fabricated microbubbles and the probing tip, as illustrated in Fig. 4(c). The small transparent cubic particles visible on the glass substrate in Fig. 4(c) correspond to the 2PP-fabricated cubic microbubbles, while the pointed probe approaches the target microbubble highlighted by the red circle. During manipulation, the microbubbles were inclined to unintended motions: they could readily rebound due to elastic forces following bond rupture or adhere to the probe as a result of electrostatic attraction, rendering them unsuitable for imaging. Additionally, forces induced by the SEM electron beam could passively displace the microbubbles once their adhesion to the substrate had been broken during the probe-tapping process, causing them to be lost within the SEM chamber.

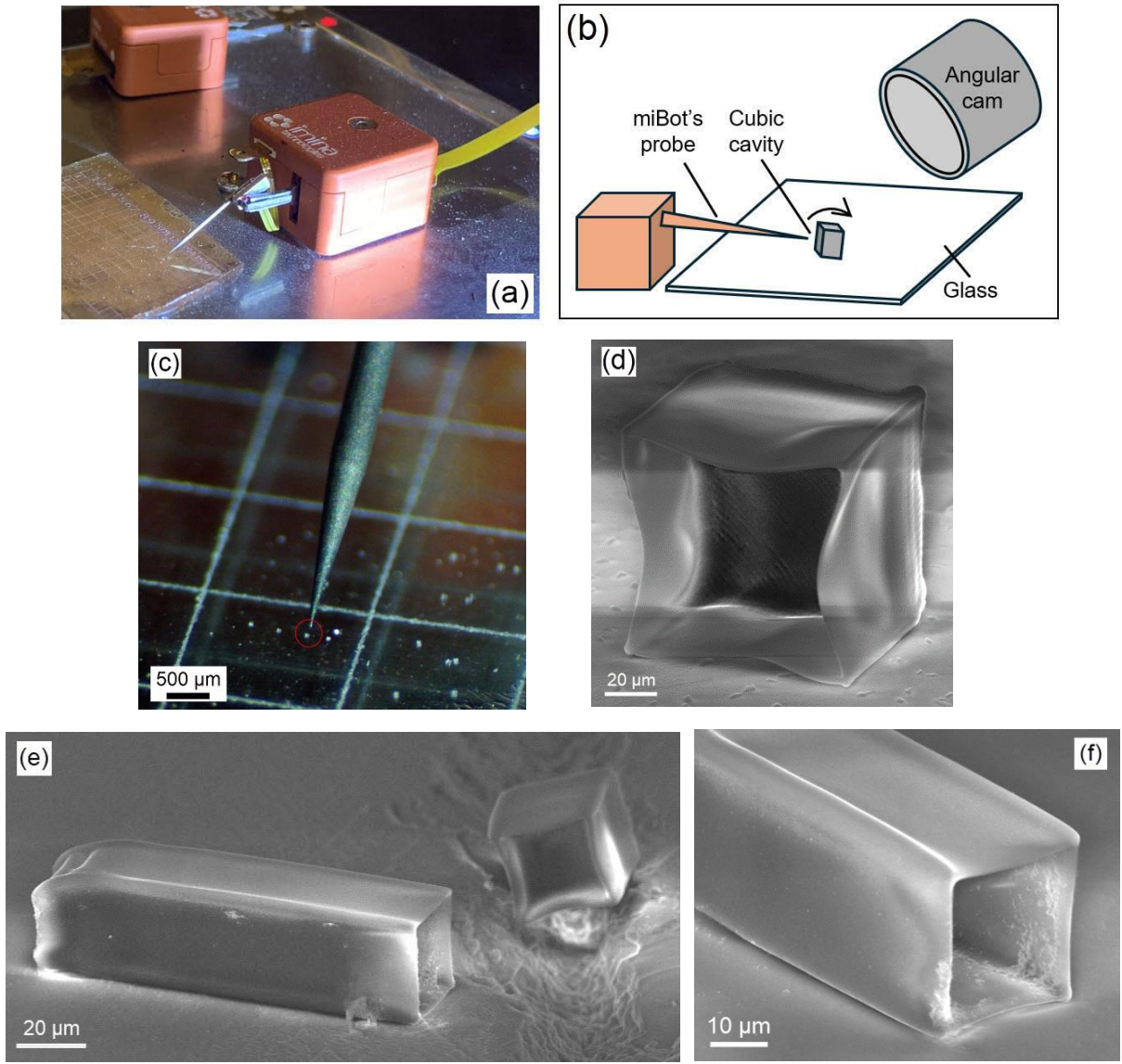

Fig. 4 (a) Piezo-actuated micromanipulator – miBot® Imina Technologies. (b) Schematic of the probe-tapping experiment used for microcavity verification. (c) Pointed probe and fabricated microbubbles imaged using an angular magnifying camera. SEM images of the hollow interiors of the tipped-over microcavities with (d) a side length and height of 90 μm and (e), (f) a high H/W aspect ratio.

Despite these challenges, SEM imaging of the cavity interior was successfully achieved, as shown in Fig. 4(d-f). The image clearly reveals the hollow interior of a tipped-over microbubble, with the overall geometry reproducing the original model design, comprising thin-walls capped by a roof. It is noted that the microbubbles exhibited greater structural integrity and mechanical rigidity than anticipated, with no structural damage observed during manipulation or imaging. This experiment conclusively confirms the hollow nature of the fabricated microbubbles. Meanwhile, the results in Fig. 4(e) and (f) indicate that deformation-induced gaps along the relatively short perimeter of the bottom bonding interface can still provide sufficient microchannels for developer fluid exchange, enabling effective removal of unexposed photoresist in this configuration. Therefore, the approach of microcavity verification validates the robustness and practicality of the proposed methodology and highlights, to the best of our knowledge, the first successful fabrication of microbubbles and fully enclosed microcavity structures using 2PP technology.

## 4. Conclusions

This study presents a comprehensive 2PP method for fabrication of cubic microbubbles with ultra-high aspect-ratio, fully enclosed, thin-walled microcavities using the high-viscosity SU-8 2050 photoresist. Key fabrication parameters were systematically optimised to address the primary technical challenges of removing unexposed photoresist from enclosed interior volumes and maintaining ultrathin walls, thereby enabling the reliable fabrication of the proposed structures with exceptional structural integrity and rigidity while avoiding common failure modes, such as inward collapse and deformation. A maximum aspect ratio of approximately 340:1 was achieved. This work significantly broadens the practical applicability of 2PP technology. The fabricated cubic microbubbles demonstrate strong potential for enabling advanced micro-structures/devices across diverse fields, including photonics, biomedical engineering, microrobotics, and microfluidics. In addition, the proposed method substantially reduces 2PP fabrication time for structures that do not require internal features by minimising the volume of laser scanning.

## Acknowledgements

The authors acknowledge financial support from the UK Royal Society International Exchanges programmes (IEC/NSFC/233421 and IES/R1/251098) and gratefully acknowledge Dr Zhipeng Chang and Dr Yanhua Hong for their valuable contributions to Java translator coding in this work.